\documentclass[%
 reprint,
 amsmath,amssymb,
 aps,
 physrev,
prb,
floatfix,
]{revtex4-2}
\pdfoutput=1

\usepackage{graphicx}
\usepackage{xcolor}
\usepackage[colorlinks=true,urlcolor=blue,
            linkcolor=blue,citecolor=blue,citebordercolor= blue,
            pdfborder={0 0 0}]{hyperref}

\begin{document}

\title{Effect of chemical disorder on the electronic stopping of solid solution alloys}
\thanks{Cite as: \href{https://doi.org/10.1016/j.actamat.2020.06.061} {Act Mater. {\bf196}, 576 (2020)}}

\author{Edwin E. Quashie}
\author{Rafi Ullah}
\email[Electronic address: ]{ullah1@llnl.gov}
\author{Xavier Andrade}
\author{Alfredo A. Correa}
\address{Quantum Simulations Group, Lawrence Livermore National Laboratory, 7000 East Avenue, Livermore, California 94550, USA}

\date{\today}

\begin{abstract}
The electronic stopping power of nickel-based equiatomic solid solutions alloys NiCr, NiFe and NiCo for protons and alpha projectiles is investigated in detail using real-time time-dependent density functional theory over a wide range of velocities.
Recently developed numerical electronic structure methods are used to probe fundamental aspects of electron-ion coupling non-perturbatively and in a fully atomistic context, capturing the effect of the atomic scale disorder.
The effects of particular electronic band structures and density of states reflect in the low velocity limit behavior.
We compare our results for the alloys with those of a pure nickel target to understand how alloying affects the electronic stopping.
We discover that NiCo and NiFe have similar stopping behavior as Ni while NiCr has an asymptotic stopping power that is more than a factor of two larger than its counterparts for velocities below \(0.1~\mathrm{a.u.}\).
We show that the low-velocity limit of electronic stopping power can be manipulated by controlling the broadening of the \(d\)-band through the chemical disorder.
In this regime, the Bragg's additive rule for the stopping of composite materials also fails for NiCr.
\end{abstract}

\maketitle

\section{Introduction}

The study of energetic charged particles shooting through materials has received great attention over many decades \cite{Rutherford_1911zz,Thomson_1912}.
Interest in understanding the underlying physics and emergent applications have been driving research in this area.
The interaction of a charged swift particle (projectile) with a target material can be quantified by the dissipative force that it experiences as it slows down, which is conventionally referred to as stopping power. 
Stopping power is defined as the energy lost per unit distance, \(S = \frac{\mathrm{d}E}{\mathrm{d}x}\) (see Ref.~\cite{Sigmund_1998} for a historical perspective on the stopping power).
From the point of view of the target material, this energy is transferred to both the host nuclei and the host electrons.
Since these two loss mechanisms are rather distinct and dominate in different velocity regimes, they are usually treated separately as \emph{nuclear} stopping power (\(S_\text{n}\)) and \emph{electronic} stopping power (\(S_\text{e}\))~\cite{Bohr_1948}. 
The nuclear stopping power is important at only very low velocities. 
The electronic stopping power is the dominant effect at high velocities or along the special channeling trajectories, in which either by coincidence or by experimental design the projectile travels long distances in crystalline directions avoiding head-on collisions with target ions~\cite{Sand_2019,Eriksson_1967}.

The accurate characterization of electronic stopping power is of critical importance to radiation damage research, with wide ranging applications in reactor engineering \cite{Granberg_2016}, space electronics \cite{Bagatin_2015}, material science \cite{Townsend_1987}, nanoscience \cite{Krasheninnikov_2010}, and medicine \cite{Levin_2005}.
It is only recently that numerical electronic structure methods are available to obtain electronic stopping power non-perturbatively and in a fully atomistic fashion~\cite{Correa_2018}, i.e., beyond historical approaches such as the jellium model, binary collision and linear response approximations. 
Thanks to these advances in the electronic structure methods, it is now possible to calculate the electronic stopping power of complex materials such as alloys and compounds, taking stock of elemental heterogeneity and consequently modified band structures which is the main subject of this work.

One of the principal goals of nuclear materials research is to minimize the effect of radiation damage.
Embrittlement and volumetric swelling are two specific quantities to minimize in structural and containment materials.
Ni content in traditional alloys has been known for its mitigating effects against swelling under irradiation~\cite{Bates_1981}. 
A new type of alloys, formally referred to as single-phase concentrated solid solution alloys (SP-CSAs) have been successfully synthesized~\cite{Yeh_2004, Cantor_2004, Gludovatz_2014}. 
Unlike traditional alloys, these equiatomic alloys are random solid solutions with a well defined underlying crystal structure, usually fcc. 
Different physical properties of these alloys such as better resistance to radiation damage, corrosion resistance, heat resistance, lower thermal expansion coefficient, good wear resistance, higher tensile strength, and higher electrical resistivity makes them a unique choice for applications in harsh conditions~\cite{Jin2017}. 

Recently, Zhang \emph{et al.}~\cite{Zhang_2015, Zhang_2016} have shown that the increase in chemical disorder, from pure Ni to equiatomic binary and quaternary solid solutions, leads to a significant reduction in electron mean free paths and thermal conductivity. 
This, in turn, causes slower heat dissipation significantly modifying defect evolution under ion irradiation. 
The overall improvement in radiation resistance is observed with increasing chemical disorder. 
The chemical disorder in these novel materials, which runs over the motif of an otherwise ordered lattice, makes the electronic structure distinct compared to both pure crystalline metals and amorphous alloys. 
For a critical review on the current trends on these alloys see Ref.~\cite{MIRACLE2017448}.

In this work, we have studied the electronic stopping power of H and He projectiles in NiCr, NiFe, and NiCo, a set of model fcc binary equiatomic random solid solution alloys, and in pure Ni and pure Cr for comparison. 
We have used real-time time-dependent density functional theory (RT-TDDFT) to compute the electronic stopping power, in particular the composition-dependency at low velocities and its relationship with the unperturbed band structure and other ground state properties.

\section{Method}

Time-dependent density functional theory (TDDFT)~\cite{Runge_1984} is a reformulation of time-dependent quantum mechanics in the same way as density functional theory (DFT)~\cite{Hohenberg_1964} is a reformulation of time-independent quantum mechanics.
Practical approximations to TDDFT can be expressed in the time-dependent Kohn-Sham (TDKS) framework~\cite{marques_2012}.
To calculate the electronic stopping power we solve numerically the time-dependent Kohn-Sham (TDKS) equations,
in atomic units (a.u.)~\cite{Kohn_1965,Quashie_2017},
\begin{equation}
    \mathrm{i} \frac{\partial\varphi_i}{\partial t}(\mathbf{r}, t) = 
        \bigg\{
            \frac{-\nabla^2}{2} 
            + v_\text{ext}(\mathbf{r}, t) + v_\text{H}[n](\mathbf{r}) 
            +
             v_\text{xc}[n](\mathbf{r})
        \bigg\}
        \varphi_i(\mathbf{r}, t),
\label{eq:tdks1}
\end{equation}
where the electronic density is given by
\begin{equation}
n(\textbf r, t) = \sum_{i}{|\varphi_i(\textbf r, t)|}^2,
\label{eq:tdks2}
\end{equation}
and \(\{\varphi_i\}\) are the single-particle (KS) electronic states,
\(v_\text{ext}\) is the time-dependent external potential (produced by the moving nuclei), 
\(v_\text{H}[n]\) is the Hartree potential that describes the classical mean-field interaction of the electron distribution 
\(n\), 
and \(v_\text{xc}[n]\) is the quantum-mechanical exchange and correlation (XC) potential which is calculated in the non-magnetic adiabatic local density approximation (LDA)~\cite{Ceperley_1980, Perdew_1981}.
Although other dynamic exchange and correlation effects that exist beyond the LDA are known to play a role in the electronic stopping of ions in jellium~\cite{Nazarov_2007}, for atomistic systems and at velocities and stopping regimes considered in this work, the corrections are expected to be small and impractical in the context of the present numerical simulation. 

The calculations are carried out by using \textsc{qbox/qb@ll}, a general purpose first principles electronic structure code~\cite{Gygi_2008}, with custom modification for time-dependence~\cite{Draeger_2016}. 
A detailed procedure of how the electronic simulations are performed and the reliability of this approach has been reported in a recent review~\cite{Correa_2018}. In particular, the electronic stopping power of pure Ni experimentally measured by Tran \emph{et. al.} (2019) \cite{tran_2019} and theoretically predicted in our previous work \cite{Quashie_2018a} using the present approach are in good agreement. 

\begin{figure}
	\begin{center}
	\includegraphics[width=0.40\textwidth]{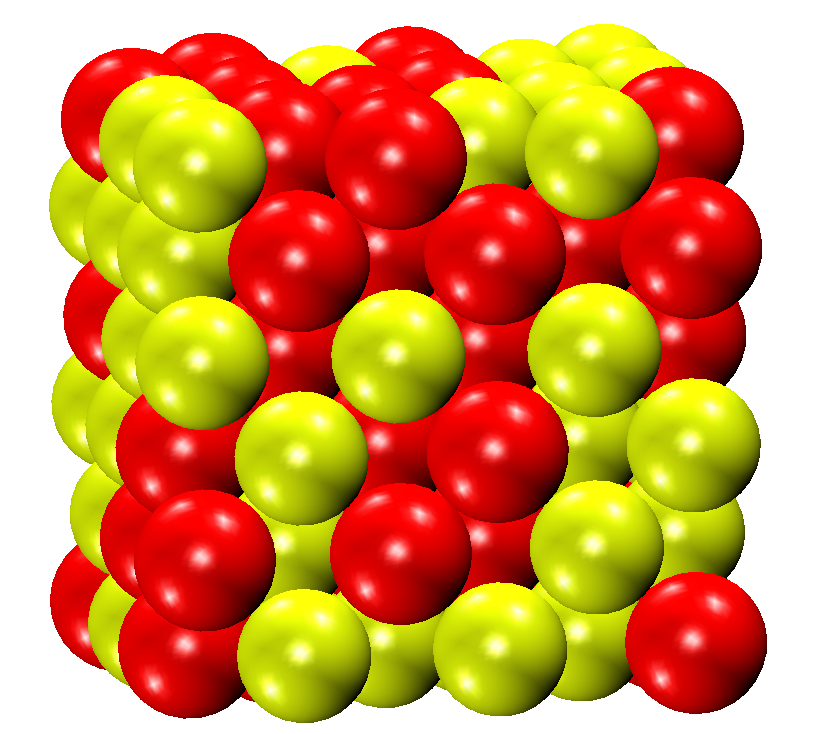}
	\caption{\label{fig:schematic_diagram_NiFe}
		A sample crystal structure of \(108\)~atoms of equiatomic fcc NiFe showing \(54\)~atoms of Ni (red) and 
		\(54\)~atoms of Fe (light green). 
		Fe is subsequently replaced by Cr and Co to obtain NiCr and NiCo in the different simulations.
	}
	\end{center}
\end{figure}

A supercell of \(108\)~atoms is constructed from \(3\times 3\times 3\) fcc conventional cubic cells using the pure Ni lattice constant of \(3.52~\mathrm{\text{\AA}}\). 
We further employ periodic boundary conditions to obtain a reasonable representation of a bulk system. 
The convergence of electronic stopping power with respect to size effects has been tested elsewhere~\cite{Correa_2018}. 
To model the random alloy, \(54\)~atoms at random sites are Ni while the other 54 random sites are replaced with Fe, Cr, or Co to get equiatomic NiFe, NiCr, and NiCo, respectively. 
The NiFe alloy in fcc structure with randomly placed Ni and Fe atoms is shown in Fig.~\ref{fig:schematic_diagram_NiFe}.
Although a slight (negative) short-range order has been found in these alloys~\cite{Tamm_2015}, we use random shuffling and the pure Ni lattice constant on a perfect (unrelaxed) fcc lattice because the densities of these random alloys structures  are very close to that of pure Ni~\cite{Jin2017}, and more importantly, because we are only interested in studying \emph{chemical} disorder effects consistently across these three alloys. 

The KS orbitals are expanded in plane-wave basis with an energy cutoff of \(160~\mathrm{Ry}\).
The ions are represented by the norm-conserving Troullier-Martins pseudopotentials~\cite{Troullier_1991} factorized in the Kleinman-Bylander form \cite{Kleinman_1982}, and only electrons in the \(3p\), \(3d\) and \(4s\) are explicitly simulated.
The convergence of electronic stopping power with respect to the size of basis set and assumptions about atomic state participation has been tested elsewhere~\cite{Ullah_2018}.

As initial condition for the real time propagation, we first calculate a self-consistent ground state of the system plus the projectile.
The time-independent wavefunctions, thus obtained, are then propagated using TDKS equations (Eq.~\ref{eq:tdks1}) while the projectile is given a finite velocity which persistently deposits energy in the system. 
The time step used in integration is about \(1\)~attosecond; while the total simulation time spans several femtoseconds. 
The atomic positions of the host atoms are fixed to restrict the dissipation to the electronic subsystem only.
The total energy of the electronic subsystems is recorded as a function of the distance traveled by the projectile. 
The slope of this function gives us the electronic stopping for that particular velocity and configuration. 
For a discussion on the definition of the energy in the context of TDKS equations see Ref.~\cite{Schleife_2015}.
It is important to note that we do not control the charge around the projectile as it is part of the dynamics.

\begin{figure}[t]
	\begin{center}
	\includegraphics[width=1.00\columnwidth]{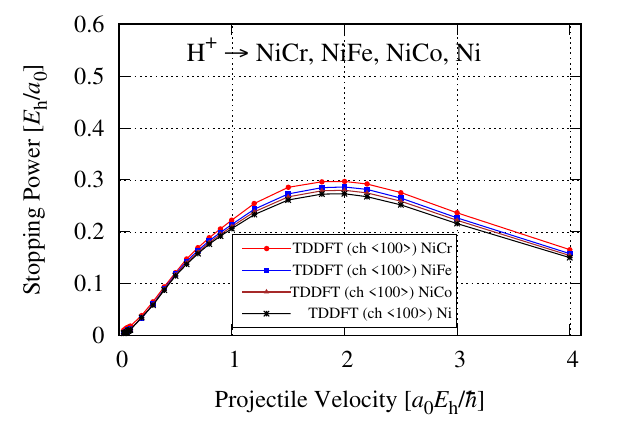}
	\caption{\label{fig:H-channel}
		For \(\mathrm{H^+}\) (proton) in \(\mathrm{NiCr}\) (red), \(\mathrm{NiFe}\) (blue), \(\mathrm{NiCo}\) (brown) and pure \(\mathrm{Ni}\) (black),
		TDDFT-simulated average electronic stopping power for the channeling (ch) geometry is compared. 
		Stopping power and velocities are in atomic units, \(E_\text h\) denotes Hartree energy unit. 
	}
	\end{center}
\end{figure}

\begin{figure}[t]
	\begin{center}
	\includegraphics[width=1.00\columnwidth]{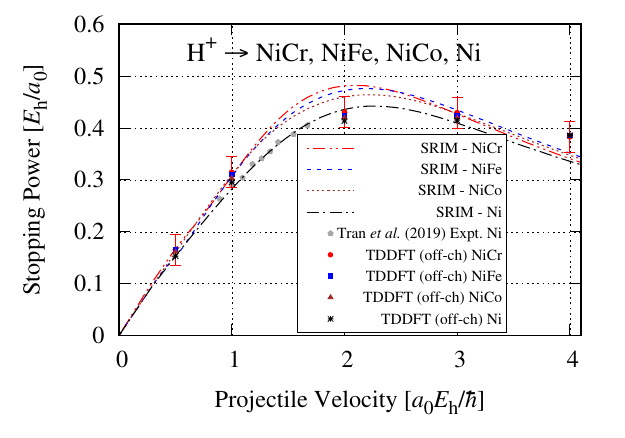}
	\caption{\label{fig:H-off-channel}
		\(\mathrm{H^+}\) (proton) in \(\mathrm{NiCr}\) (red), \(\mathrm{NiFe}\) (blue), \(\mathrm{NiCo}\) (brown) and pure \(\mathrm{Ni}\) (black),
		TDDFT-simulated average electronic stopping power (symbols-only) is compared with the SRIM model~\cite{Ziegler_2010} (dashed) for off-channeling (off-ch) random directions and recent experimental results for pure {Ni} \cite{tran_2019} (for additional comparisons with experiment see Ref. \cite{Quashie_2018a}). 
		Vertical bars represent typical errors from random sampling.
		Stopping power and velocities are in atomic units, \(E_\text h\) denotes Hartree energy unit. 
	}
	\end{center}
\end{figure}

Although long known, the contributions of the core electrons in the host atoms to the electronic stopping power have recently been quantified by first principles simulations~\cite{Ullah_2018}. 
We have explicitly treated the semicore electrons in the pseudopotential approximation to take into account possible additional energy dissipation. 
The pseudopotential representing Ni has 16 valence (explicitly simulated) electrons (\(3p^{6}3d^{8}4s^{2}\)), Cr has 12 electrons (\(3p^{6}3d^{5}4s^{1}\)), Fe has 14 electrons (\(3p^{6}3d^{6}4s^{2}\)), and Co has 15 electrons (\(3p^{6}3d^{7}4s^{2}\)).

We considered two distinct simulation setups in this study.
First, 
	we considered the \emph{channeling} case where the projectile moves in a straight line avoiding collisions with the target atoms, and 
secondly, 
	the \emph{off-channeling} case where the projectile moves in a randomly chosen direction in the crystal to probe larger electronic density around the target atoms. 
The channeling trajectory considered in this work, in which the projectiles are shot at the center of the channel along the \(\langle 100\rangle\) direction with zero angle of incidence, is an exceptional preparation in real experiments.
On the other hand, the off-channeling type of trajectory mimics most experimental settings (and empirical models such as SRIM's~\cite{Ziegler_2010}), allowing a direct comparison, and is expected to yield larger stopping values. 
The sampling of random trajectories is described in Refs.~\cite{Quashie_2016,Quashie_2018a}.

\section{Results}

The TDDFT simulation results for protons in NiCr, NiFe, NiCo and Ni are compared for the channeling geometry as shown in Fig.~\ref{fig:H-channel}.
The off-channeling results are shown in Fig.~\ref{fig:H-off-channel} and compared with the empirical SRIM model~\cite{Ziegler_2010} of electronic stopping power in a wide range of proton velocities \(0.04 \le v \le 4.0~\mathrm{a.u.} \) (\(40~\mathrm{eV} - 400~\mathrm{keV}\)). 
SRIM is based on a phenomenological method which uses a combination of models, fitting parameters, density-scaling and additive chemical rules to produce stopping curves for arbitrary compounds and projectiles. 
In the absence of direct experimental data, it is the best available empirical estimate of electronic stopping power. 
It is worth mentioning that due to current experimental limitations the electronic stopping power cannot be measured for projectile velocities \(v \lesssim 0.1~\mathrm{a.u.}\) (or energies \(\lesssim 250~\mathrm{eV}\)) \cite{Roth_2017a}. 
Thus, any empirical data below this velocity limit is not reliable and any comparison with it should be deemed as such. 
The size effects and need for harder potentials set an upper limit in terms of the projectile velocity up to which these numerical simulations can provide a reliable estimate of the electronic stopping power; 
results for large velocities (above \(4~\mathrm{a.u.}\) are not reported here.
These limitations could be overcome, but only at an exuberant computational cost often beyond the reach of available high performance computing platforms.

The uncertainty in channeling data is dominated by the slope determination from the energy uptake curves and does not depend on the choice of channel or impact parameter. 
This work is limited to hyperchanneling (center of the channel) in the \(\langle 100\rangle\) channel only, a representative of the channelling condition, likely with the lowest effective stopping power.
The uncertainly in the off-channeling calculations is not only comparatively larger but is expected to strongly depend on the choice of geometry. 
The trajectories in our off-channeling setup are chosen such that the projectile traverses a distance equivalent to several lattice parameters, providing an effective averaging over the varying chemical environment.
Another source of uncertainty in the theoretical calculations is the choice of different random structures.
An accurate estimate of such an error requires many additional calculations and is beyond the scope of the current work.

In the simulations \(S_\text{e}\) in the \(\langle 100\rangle\) channel is \(35\%\) smaller than the SRIM prediction at \(v \sim 1~\mathrm{a.u.}\). 
This difference when comparing channeling and off-channeling \(S_\text{e}\) is expected and consistent with previous simulation works~\cite{Quashie_2016,Quashie_2018a}. 
The result for NiCr shows a slightly higher \(S_\text{e}\) when compared with its counterparts (NiFe, NiCo) and Ni across the overall range both for the channeling and off-channeling case, although NiCr has fewer number of valence electrons in the simulation.

\begin{figure}
	\begin{center}
		\includegraphics[width=1.00\columnwidth]{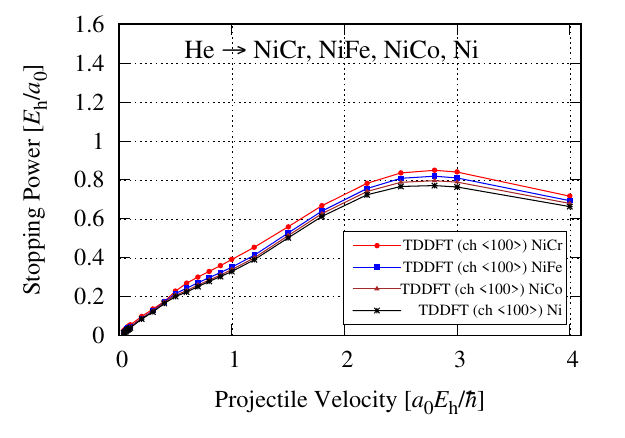}
		\caption{\label{fig:He-channel}
			He in \(\mathrm{NiCr}\) (red), \(\mathrm{NiFe}\) (blue), \(\mathrm{NiCo}\) (brown) and pure \(\mathrm{Ni}\) (black),
			TDDFT-simulated average electronic stopping power for a channeling (ch) \(\langle 100\rangle\) geometry.
		}
	\end{center}
\end{figure}

Moving on to the off-channeling case, our simulation results are in good agreement with the empirical model data for velocities in the simulated range \(0.5 - 3.0~\mathrm{a.u.}\). 
For the SRIM model there is a shallow cross-over where the \(S_\text{e}\) for NiCr is smaller than that of NiFe and NiCo above \(v = 2.4~\mathrm{a.u.}\). 
Although it is a slight effect, marginally over the statistical error of random direction sampling, this already shows the limitations of assuming that electrons form a homogeneous electron gas (jellium), since in the atomistic context not all valence electrons participate equally.
We attribute this to the different electronic structures of the alloys, for example as described by the density of states (DOS) of the targets, which we discuss in detail later in this section as the trend is more marked at lower velocities.

\begin{figure}
	\begin{center}
		\includegraphics[width=1.00\columnwidth]{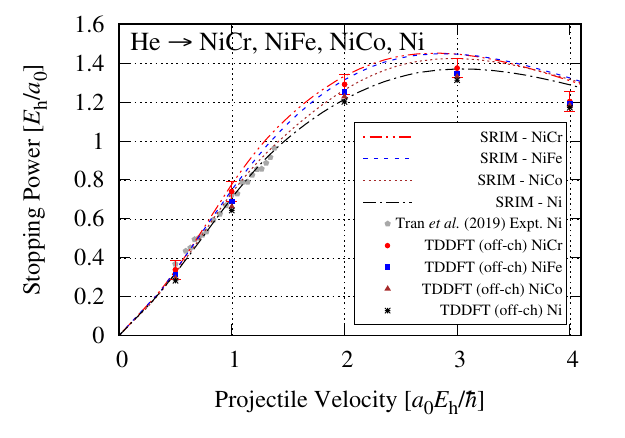}
		\caption{\label{fig:He-off-channel}
			He in \(\mathrm{NiCr}\) (red), \(\mathrm{NiFe}\) (blue), \(\mathrm{NiCo}\) (brown) and pure \(\mathrm{Ni}\) (black),
			TDDFT-simulated average electronic stopping power (symbols-only) is compared with the SRIM model~\cite{Ziegler_2010} (dashed) for off-channeling (off-ch) random directions and with recent experimental results for pure Ni \cite{tran_2019} (for additional comparisons with experiment see Ref. \cite{Quashie_2018a}).
			Vertical bars represent typical errors from random sampling (only shown for NiCr).
		}
	\end{center}
\end{figure}

The simulation results for \(\mathrm{He}\) projectile in the same set of targets are shown in Figs.~\ref{fig:He-channel} and \ref{fig:He-off-channel}.
Again, the simulated channeling \(S_\text{e}\) is significantly lower than the empirical stopping, and there is a good agreement between off-channeling results with the SRIM data for most of the velocity range considered (\(0.5-4~\mathrm{a.u.}\)). 
For these velocities, simulation results consistently show that NiCr has the highest \(S_\text{e}\), typically followed by NiFe, NiCo, and Ni, albeit the difference is small. 
For both projectiles, the calculated \(S_\text{e}\) in Ni remains the lowest compared to the alloys for all velocities.

The maximum of \(S_\text{e}\) for proton and He in the NiCr target is \(0.482~E_\text h/a_\text 0\) and \(1.452~E_\text h/a_\text 0\), respectively. 
According to linear response theory \cite{Lindhard_1964_book}, the \(S_\text{e}\) quadratically depends on the projectile ion's charge \(Z_1\), for a given electronic medium target.
This means that the difference in \(S_\text e\) between the two fully stripped projectiles, in this case a proton and an alpha particle, should be a factor of \(4\).
Instead, in our non-linear calculations we find a factor of \(\sim 3\) near the maximum, which can be attributed to a partial ionization of the alpha projectile; for example \(Z^*_1 = \sqrt{3} = 1.73\) while assuming a similar host electron participation and a fully stripped proton.
This analysis of partial ionization is characteristic of alpha particles in these contexts~\cite{Echenique_1986};
for example, a more careful analysis carried out in Ref.~\cite{Quashie_2018a} in pure Ni shows a value of \(Z^*_1 = 1.5\) for alpha projectiles in pure Ni at \(v \sim 2~\mathrm{a.u.}\)

\begin{figure}[h!]
	\begin{center}
	\includegraphics[width=1.00\columnwidth]{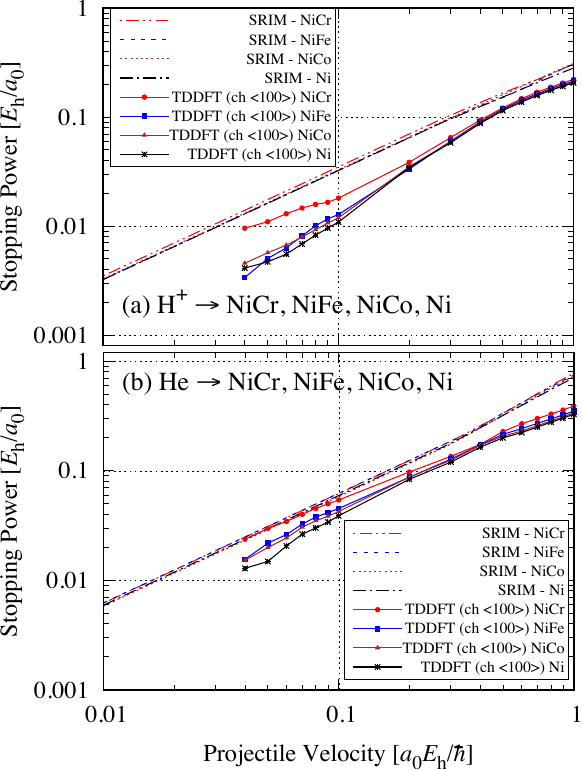}
	\caption{
		\(\mathrm{H^+}\) (top panel) and \(\mathrm{He}\) (bottom panel) in~\(\mathrm{NiCr, NiCo, NiFe} \) and~\(\mathrm{Ni}\). 
		The average electronic stopping power for the projectiles versus projectile velocity. 
		The solid curves indicate results for TDDFT in a channeling (ch) \(\langle100\rangle\) direction.
		The dashed lines refer to results obtained from SRIM database~\protect\cite{Ziegler_2010}.
		\label{985007}
	}
	\end{center}
\end{figure}

Apart from the quantitative trends mentioned so far, there is no qualitative difference between the alloys or between the alloys and the pure Ni for high velocity projectiles (\(v > 0.5~\mathrm{a.u.}\)).
Their corresponding \(S_\text e\) simply shows a slight scaling factor and the differences are marginally close to the error bar inherent to the simulation method for off-channeling trajectories. 
However, at low velocities the situation might be different.
The electronic band structure is known to have dramatic differences near the Fermi level for these alloys.
It has been hypothesized that these differences in the energy scale are directly responsible for the unique properties of these alloys regarding to energy transport, and indirectly to radiation resistance~\cite{Zhang_2015, Zhang_2016}.
At low velocities it is more practical and representative to consider channeling trajectories.
The channeling case for a given low index direction has much smaller statistical error bars because they mainly depend on length limitations of the trajectory (simulation cell), allowing a reliable analysis at low velocities. 
The low velocity (\(0.04-1.0~\mathrm{a.u.}\)) \(S_\text e\) data are replotted on the logarithmic scale and are shown in Fig.~\ref{985007}~(a) and \ref{985007}~(b). 
In the case of the proton projectile, we observe that the \(S_\text{e}\) is very similar for NiCo, NiFe and Ni with less than \(10\%\) difference for velocities below \(0.1~\mathrm{a.u.}\), as shown in Fig.~\ref{985007}(a).
In the NiCr case, we see a significant difference up to a factor 3 compared to the other alloys, with a clear separation from the other cases. 
A similar trend is seen for helium in NiCr where the \(S_\text e\) is higher compared to its counterparts for \(v < 0.4~\mathrm{a.u.}\), as shown in Fig.~\ref{985007}(b).

\begin{figure}
	\begin{center}
	\includegraphics[width=1.00\columnwidth]{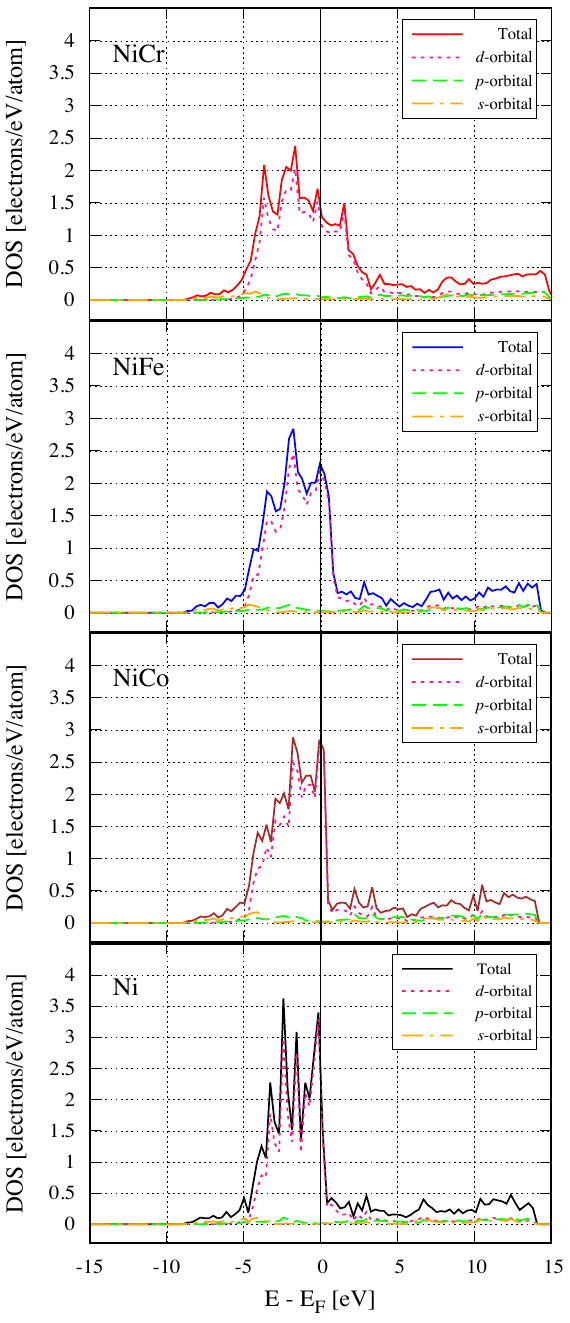}
	\caption{
		The electronic densities of states (DOS) are shown for~\(\mathrm{NiCr}\),~\(\mathrm{NiFe}\),~\(\mathrm{NiCo}\) and~\(\mathrm{Ni}\) as function of~\(E-E_\text F\). 
	(The vertical black line corresponds to the Fermi energy \(E_\text F\)). 
	Area below and above the Fermi energy corresponds to the occupied and unoccupied states respectively in the pristine material. 
		\label{fig:246853}
	}
	\end{center}
\end{figure}

In order to understand the features of \(S_\text e\) for the alloys at low velocities, we calculated their ground-state electronic structure by means of DFT as implemented in the VASP code~\cite{Kresse_1996,Kresse_1996a}. 
We have used the same 108 atom supercells with no projectile.
We used the non-magnetic LDA exchange-correlation functional and a planewave energy cutoff consistent with that of the TDDFT simulations for the NiCr, NiFe, NiCo and Ni systems.
The projected augmented wave (PAW) potentials~\cite{Blochl_1994,Kresse_1999} were utilized for the core-valence electrons interaction.
The Brillouin zone is sampled by a \(6\times 6\times 6\) Monkhorst-Pack grid of k-points~\cite{Monkhorst_1976} and in evaluating the electronic density of states (DOS), the occupancies of the electronic states are determined with the tetrahedron method.

The unperturbed DOS of the three alloys and Ni are shown in Fig.~\ref{fig:246853}, where the Fermi energy \(E_\text F\) is at zero. 
The DOS at \(E \simeq E_\text F\) is dominated by the \(d\)-electrons for NiCr, NiFe, NiCo and Ni. 
The \(d\)-band for NiCr is broader compared with NiFe, NiCo and Ni.
At \(E \approx E_\text F\), there are no significant differences in the \(s\) and \(p\) states for the targets. 
The DOS integrated from \(-15~\mathrm{eV}\) up to \(E_\text F\) yields values of \(8\), \(9\), \(9.5\) and \(10\) electrons per atom for NiCr, NiFe, NiCo and Ni respectively; i.e., the number of simulated valence electrons likely to participate to some degree in the extreme low velocity electronic stopping.
These differences are not by themselves enough to explain the difference in behavior, because they are small and also do not strictly follow the pattern in the relative magnitude ordering of \(S_\text e\) (see Fig.~\ref{985007}).
From previous work~\cite{Quashie_2016, Quashie_2018a}, we know that for the Period \(4\) transition metals, projectiles at velocities below \(0.8~\mathrm{a.u.}\), for example, depend critically on the electronic structure in a range of \(\sim 3.2~\mathrm{eV}\) around the Fermi energy (according to the estimate \(E_\text F \pm 2 k_\text F v\) derived in Ref.~\cite{Quashie_2016}).

From this insight, we can qualitatively explain the behavior of \(S_\text e\) for the target species.
NiCr has higher electronic DOS below \emph{and} above \(E_\text F\), therefore allows for low energy excitations of electrons close to the Fermi surface with higher probabilities~\cite{Roth_2017, Roth_2017a}. 
This is responsible for the relatively high \(S_\text e\) observed for NiCr at low velocities (see Fig.~\ref{985007}) for both \(\mathrm{H}\) and \(\mathrm{He}\) projectiles;
while low DOS above the Fermi level for NiFe, NiCo and Ni is responsible for the low \(S_\text e\) compared to that of the NiCr target. 
This behavior is expected to become more pronounced at even lower velocities not accesible by our direct simulation method.
Similar behavior was observed for transition and rare earth metals by Roth~\emph{et al.}~\cite{Roth_2017}.

\begin{figure}
	\begin{center}
	\includegraphics[width=1.00\columnwidth]{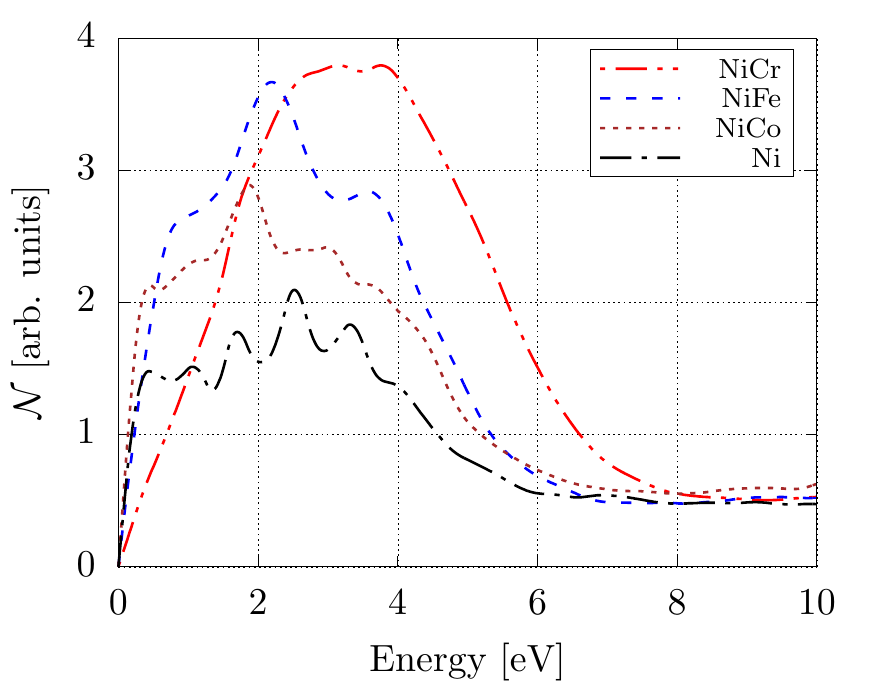}
	\caption{
		Number of available transitions with energy \(\hbar\omega\) from occupied to unoccupied states within the \(d\)-band in the alloys \(\mathrm{NiCo}\), \(\mathrm{NiFe}\) and \(\mathrm{NiCr}\) and pure \(\mathrm{Ni}\). 
		\(\mathrm{NiCr}\) shows the largest availability of transitions in the range of energies above \(2~\mathrm{eV}\).
		This availability of states correlates with the electronic stopping power result below \(v = 0.2~\mathrm{a.u.}\) (Fig.~\ref{985007}).
		\label{fig:cdos}
	}
	\end{center}
\end{figure}

This argument can be quantified by analysis such as the joint-density of states~\cite{marder_2010}. 
Given the non-dispersive nature of the \(d\)-bands and the fact that there is no strict crystalline momentum conservation~\cite{Zhang_2015,Zhao_2018} 
due to the chemical disorder in these random alloys, a simpler analysis can be done via a conditional density of states, defined simply as,
\begin{equation}
 \mathcal{N}(\hbar\omega) = \int_0^{\hbar\omega}\mathcal{D}_\mathrm{occ}(E - E_\text F - \hbar\omega)\mathcal{D}_\mathrm{unocc}(E - E_\text F) \mathrm{d}E,
\end{equation}
where \(\mathcal{D}\) is the DOS (the left/right factor always evaluated for occupied/unoccupied states). 
This conditional density of states counts the number of available occupied to unoccupied possible transitions in an energy range \(\hbar\omega\).
Without taking into account selection rules like conservation of crystalline momentum of the projectile-electron collisions~\cite{Artacho_2007,Ullah_2015}, this is the simplest analysis we can make in these random crystals where there is no strict concept of a unit cell Brillouin zone or energy-momentum dispersion~\cite{Zhang_2015,Zhao_2018}. 
By this analysis it is clear that the main difference between \(\mathrm{NiCr}\) and the rest of the targets is that it has a larger number of possible transitions in the energy range above \(3~\mathrm{eV}\), which explains the consistent qualitative difference for \(v < 0.1~\mathrm{a.u.}\) (see Fig.~\ref{fig:cdos}).

\begin{figure}[t]
	\begin{center}
	\includegraphics[width=1.00\columnwidth]{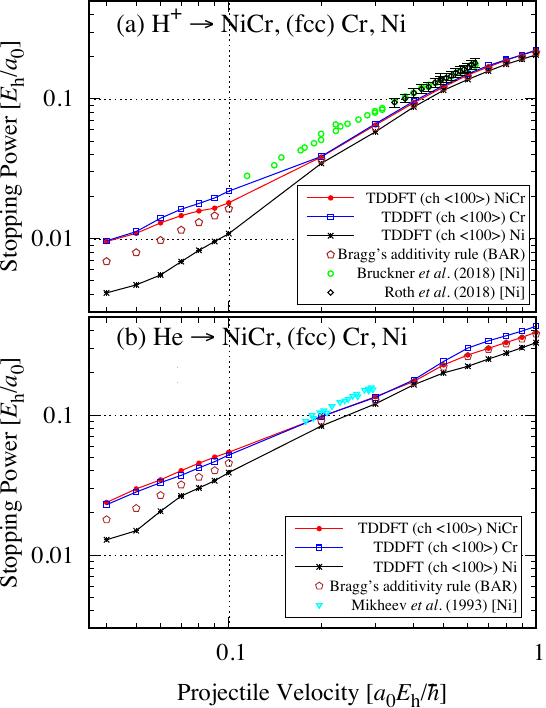}
	\caption{
		\(\mathrm{H^+}\) (top panel) and \(\mathrm{He}\) (bottom panel) in \(\mathrm{NiCr, Cr}\) and \(\mathrm{Ni}\). 
		The \(S_\text{e}\) for the projectiles versus projectile velocity~\(v\). 
		The solid curves indicate results for TDDFT channeling (ch) directions.
		The brown hexagonal points indicates BAR applied to the individual simulation results of \(\mathrm{Ni}\) and (fcc) 
		\(\mathrm{Cr}\) while the circles show the results of the explicit simulation with NiCr.
		\label{445143}
	}
	\end{center}
\end{figure}

From the preceding analysis it is feasible that the low-velocity limit \(S_\text e\) can be manipulated by controlling the broadening of the \(d\)-band.
Looking at the unperturbed DOS for the three alloys and Ni (as shown in Fig.~\ref{fig:246853}) and the electronic configurations of Ni, Co, Fe, and Cr, it appears that Ni, which has 8 electrons in the \(d\)-band when combined with Cr which has 5 electrons in its \(d\)-band, produces the largest broadening of the the resulting \(d\)-band (NiCr). 
Based on this observation we speculate that combing the first row transition metals with fewest \(d\) electrons (such as Vanadium and Titanium) with Ni may increase the low-velocity limit stopping even further; although we are not aware if these alloys have been synthesized or stable as random alloys.
NiTi (Nitinol 55/60), however, is a well known shape memory austenite alloy~\cite{Buehler1963}.
This adds to a family of known effects controlled by the \(d\)-electrons in the nickel-based alloys, such as thermophysical~\cite{Jin2017} and defect evolution phenomena~\cite{Zhao_2018,ZHAO2020}.

%%%%%%%%%%%%%%%%%%%%%%%%%%%%%%%%%%%%%
%%%%%%%%%%%%%%% Bragg rule.   %%%%%%%%%%%%%%%
%%%%%%%%%%%%%%%%%%%%%%%%%%%%%%%%%%%%%%

A detailed atomistic simulation like the one presented here allows to test techniques and approximations, such as the Bragg's additive rule, historically applied to electronic stopping power in composite materials.
The Bragg's additive rule (BAR)~\cite{Bragg_1905} is commonly used to obtain compound stopping solely from models or experimental data available for pure component systems.
The BAR approximation is in general valid at higher projectile energies, but it is eventually known to fail at lower projectile energies~\cite{Sigmund_2002}.

Given distinguishing features of NiCr found here, we apply the Bragg's additive rule (BAR)~\cite{Bragg_1905} to compare our calculated electronic stopping for NiCr with that of Ni and Cr targets. 
The original BAR postulated that the ``loss of range" of \(\alpha\) particles in a material was proportional to the weighted sum of square roots of the atomic weights of the constituent atoms relative to air. However, the modern formulation of BAR is given in terms of stopping cross section, \(\varepsilon\), and \(\varepsilon = \frac{1}{N}S\), where \(N\) is the number of atoms, molecules, or formula units per unit volume \cite{Thwaites_1983}. 
Thus, the approximation reads:

\begin{equation}
\varepsilon^{\text{compound}} = \sum_i n^{(i)} \varepsilon^{(i)},
\end{equation}

where \(n^{(i)}\) is the number of atoms of the \(i^{\mathrm{th}}\) element per molecule or formula unit.
We recast it in terms of the stopping power, the approximation reads:
\begin{equation}
S^{\text{compound}} = N^{\text{compound}} \sum_i n^{(i)} \frac{S^{(i)}}{N^{(i)}}.
\label{eq:BAR} 
\end{equation}

In atomistic simulations, in addition to the atomic density, we can exactly control the atomic structure even if it corresponds to artificial or unstable phases.
For example, since we are interested in testing BAR rule under ideal mixing, we calculated the electronic stopping in fcc pure Cr (at the same atomic density as pure fcc Ni), despite the fact that the most stable structure of Cr is not fcc, but bcc. 
We apply BAR to our channeling results for Ni and Cr and compare it to the channeling results for their alloy (NiCr), all three have the same underlying lattice structure and lattice constant. 
The BAR for our particular case where \(N^\mathrm{NiCr}= 2\) per average unit cell, \(N^\mathrm{Ni} = N^\mathrm{Cr} = 4\) per unit cell and \(n^\mathrm{Ni} = n^\mathrm{Cr} = 1\), is simply:

\begin{equation}
S_\text{e}^\mathrm{NiCr} = \frac{1}{2}S_e^\mathrm{Ni} + \frac{1}{2}S_e^\mathrm{Cr}.
\end{equation}

We observe significant deviations from BAR (up to \(30\%\)) for proton at \(v < 0.1~\mathrm{a.u.}\) but at higher velocities above \(0.3~\mathrm{a.u.}\) the validity of BAR improves as shown in Fig.~\ref{445143}(a).
This is expected since BAR fails at lower energies and agrees well at higher projectile energies~\cite{Thwaites_1983, Thwaites_1992, Golser_1992, Bauer_2004, Bar_2018}. 
A similar good agreement at higher velocities and disagreement at lower velocities for the BAR approximation have been reported in Refs.~\cite{Sigmund_2003,Sigmund_2004,Sharma_2004}.

We see similar failure of the BAR for the \(\mathrm{He}\) projectile as shown in Fig.~\ref{445143}(b). We compare our TDDFT channeling results for \(\mathrm{He}\) particles in NiCr, Ni and Cr with available experimental data and with BAR results for Ni and Cr. 
For both projectiles, the \(S_\text e\) of NiCr asymptotically approaches that of (fcc) Cr, rather than to that of Ni or the \(\mathrm{Ni+Cr}\) BAR prediction at low velocities.
Our pure Ni results agrees well with recent experimental results from Refs.~\cite{Bruckner_2018,Roth_2018} (for protons) and Ref.~\cite{Mikheev_1993} (for helium).

\section{Conclusion}

In summary, we presented first principles calculations of electronic stopping power in fcc concentrated solid solution alloys, namely, NiCr, NiFe and NiCo; and pure Ni for H and He particles.
We have shown that \(S_\text e\) for Ni-based alloys is generally higher than that of pure Ni.
At higher velocities, the stopping power is quite similar for all alloys and Ni for both channeling and off-channeling setups. 
This shows that at high projectile velocities, there is no significant difference in the type of target.

At lower velocities the stopping power in NiCr is particularly higher than those of NiFe and NiCo.
We attribute this distinct behavior to the abundance of possible transitions below and above the Fermi energy.
More electrons can be excited at lower projectile velocities compared with NiFe and NiCo, which have fewer unoccupied states by which electrons could absorb energy from the incoming projectile.
NiCr is expected to stop light ions more efficiently at low kinetic energies.
The Bragg's additive rule breaks down for NiCr below velocities of \(0.2~\mathrm{a.u.}\) at the same point where band structure effects start being important and is recovered at higher velocities.
Chemical disorder \emph{per se} does not seem to be a necessary ingredient in the explanation of the phenomena described here, except indirectly through the resulting density of states in this range of energies. 
This does not rule out the effect of disorder in the meV (electron-phonon regime) which is beyond the scope of the technique described here.

The chemical disorder does not appear to significantly affect the electronic stopping power. Therefore, the changes in radiation damage processes in SP-CSAs, particularly increased radiation resistance cannot be explained in terms of the electronic stopping power.

\section{Acknowledgements}

This work was performed under the auspices of the U.S. Department of Energy by Lawrence Livermore National Laboratory under Contract DE-AC52-07NA27344. 
The work was supported by the U.S. Department of Energy, Office of Science, Materials Sciences and Engineering Division.
Computing support for this work came from the Lawrence Livermore National Laboratory Institutional Computing Grand Challenge program.

\bibliography{Ni-alloys-stopping}

\end{document}